\documentclass[12pt,a4paper]{conference}

\usepackage{fancyhdr}
\usepackage{graphicx,amsmath,amssymb,cite}
\usepackage{multind}
\makeindex{author} \makeindex{subject}

\newcommand{\beq}{\begin{equation}}
\newcommand{\eeq}[1]{\label{#1}\end{equation}}
\newcommand{\eeqn}{\end{equation}}

\newcommand{\beqa}{\begin{eqnarray}}
\newcommand{\eeqa}[1]{\label{#1}\end{eqnarray}}
\newcommand{\eeqan}{\end{eqnarray}}

\let\bar=\overbar

\newcommand{\Dslash}{\not{\hbox{\kern-4pt $D$}}}
\newcommand{\dslash}{\not{\hbox{\kern-2pt $\del$}}}

\newcommand{\msb}{{\bar{\ssstyle M \kern -1pt S}}}

\begin{document}
%%%%%%%%%%%%%%%%%%%%%%%%%%%%%%%%%%%%%%%%%%%%%%%%%%%%%%%%%%%%%%%%%%%%%%%

\Chapter{$D$ and $\bar{D}$ mesons in hot and dense matter}
           {$D$ and $\overline{D}$ mesons in hot dense matter}{L. Tol\'os {\it et al.}}
\vspace{-5.3cm}\includegraphics[width=6 cm]{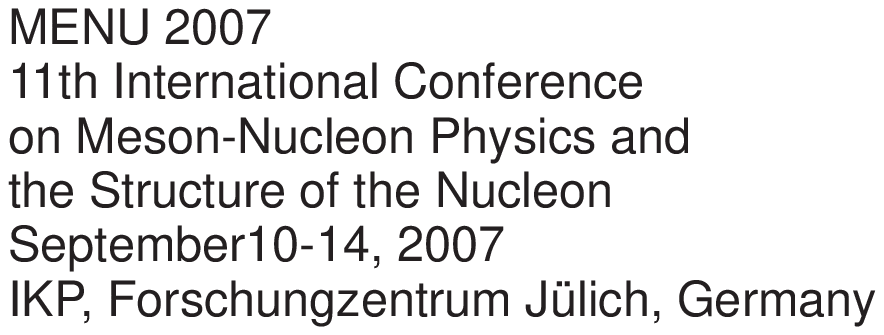}
%\bigskip\bigskip
\vspace{3cm}

\addcontentsline{toc}{chapter}{{\it N. Author}} \label{authorStart}
%%%%%%%%%%%%%%%%%%%%%%%%%%%% NEW SWITCHES %%%%%%%%%%%%%%%%%%%%%%%%%%%%%%

\begin{raggedright}

{\it L. Tol\'os$^*$ \footnote{tolos@fias.uni-frankfurt.de}, A. Ramos$^{\dag}$ and T. Mizutani$^{\S}$ }\index{author}{Tol\'os, L.}

\bigskip

$^*$FIAS. J.W.Goethe-Universit\"at. \\
Ruth-Moufang-Str. 1, 60438 Frankfurt (M), Germany

\bigskip

$^{\dag}$Dept. d'Estructura i Constituents de la Mat\`eria. Universitat de Barcelona.
Diagonal 647, 08028 Barcelona, Spain

\bigskip

$^{\S}$Department of Physics, Virginia Polytechnic Institute and State
University.\\
Blacksburg, VA 24061, USA

\end{raggedright}

\begin{center}
\textbf{Abstract}
\end{center}

The $D$ and $\bar {D}$ mesons are studied 
in hot dense matter within a self-consistent
coupled-channel approach taking, as bare interaction, a broken SU(4) $s$-wave Tomozawa-Weinberg interaction supplemented by an attractive isoscalar-scalar term. The in-medium solution at finite temperature incorporates Pauli
blocking effects, baryon mean-field bindings, and $\pi$ and open-charm meson self-energies. In the $DN$ sector, the $\Lambda_c$ and $\Sigma_c$
resonances remain close to their free-space position while acquiring a
remarkable width. As a result, the
$D$ meson spectral density shows a single pronounced peak close to the free mass that broadens with increasing density specially towards lower energies. The
low-density theorem is not a good approximation for the repulsive
$\bar D$ self-energy close to saturation density.
We discuss the implications for the $J/\Psi$ suppression at CBM (FAIR).

\vspace{0.5cm}

 The future CBM (Compressed Baryon Matter) experiment of the FAIR project at GSI aims at investigating, among others, the possible modifications of the properties of open ($D$ and $\bar D$) and hidden (e.g. $J/\Psi$) charmed mesons in a hot and dense baryonic environment. 
%%%%%%%%%%%%%%%%%%%%%%%%%%%%%%%%%%%%%%%%%%%%%%%%%%%%%%%%%%%%%%%%%%%%%%%%%%%%5
\begin{figure}[htb]
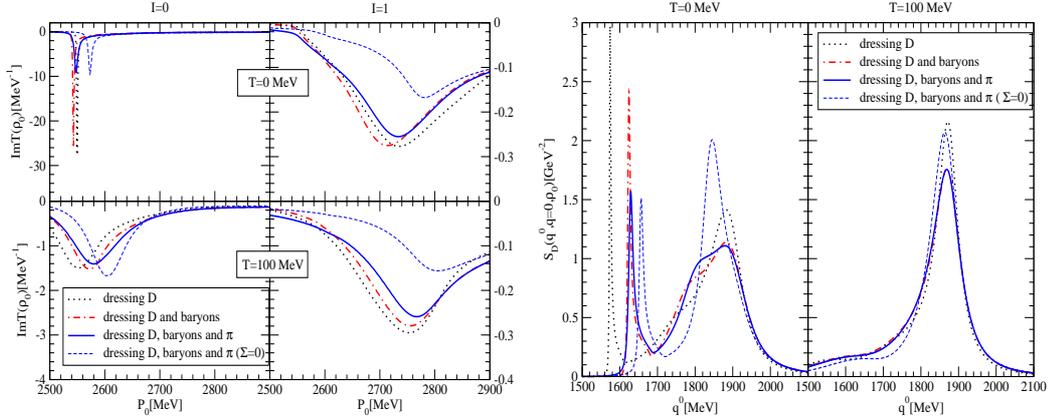

\begin{center}
\includegraphics[height=5.5 cm, width=6.7 cm]{tolos_fig2.eps}
\hfill
\includegraphics[height=5.5 cm, width=6.7 cm]{tolos_fig3.eps}
\caption{$\Lambda_c$ and $\Sigma_c$ resonances, and the $D$ meson spectral function} \label{fig:DN}
\end{center}
\end{figure}
%%%%%%%%%%%%%%%%%%%%%%%%%%%%%%%%%%%%%%%%%%%%%%%%%%%%%%%%%%%%%%%%%%%%%%%%%%%%%

The in-medium modification of the $D (\bar D)$ mesons may explain the $J/\Psi$ suppression \cite{MAT86} in an hadronic environment,  based on the mass reduction of $D (\bar D)$ in the nuclear medium \cite{TSU99}. However, a coupled-channel meson-baryon
scattering in nuclear medium is needed due to the strong coupling among the
$DN$ and other meson-baryon channels \cite{TOL04,TOL06,LUT06,MIZ06}. In the present article, we pursue a coupled-channel study on the spectral properties of $D$ and $\bar D$ mesons in nuclear matter at finite temperatures by extending the result of Ref.~\cite{MIZ06} in order to examine the possible implications in the $J/\Psi$ suppresion at FAIR. The $D$ and $\bar D$ self-energies at finite temperature are obtained by performing a self-consistent coupled-channel calculation taking, as bare interaction, a type of broken SU(4) $s$-wave Tomozawa-Weinberg (TW) interaction supplemented by an attractive isoscalar-scalar term ($\Sigma_{DN}$). The multi-channel transition matrix $T$ is solved using a cutoff regularization, which is fixed by reproducing the position and the width of the $I=0$ $\Lambda_c(2593)$ resonance while a new resonance in $I=1$ channel $\Sigma_c(2880)$ is generated \cite{MIZ06}.

The in-medium solution at finite temperature incorporates Pauli blocking effects, baryon mean-field bindings via a temperature-dependent $\sigma -\omega$ model, and $\pi$ and open-charm meson self-energies in the intermediate propagators (see \cite{TOL07}). The in-medium self-energy and corresponding spectral density are obtained self-consistently summing $T_{DN}$ over the nucleon Fermi distribution.

The behavior of the $I=0$
$\Lambda_c$ and $I=1$ $\Sigma_c$ resonances in hot dense matter is shown in the l.h.s. of Fig.~\ref{fig:DN} for three different self-consistent calculations:  i) including only the
self-consistent dressing of the $D$ meson, ii) adding the mean-field binding of baryons (MFB) and iii) 
including MFB and the pion self-energy (PD). The thick lines correspond to model A (viz.
$\Sigma_{DN} \neq 0$) while the thin-dashed lines refer to Case (iii) within model B ($\Sigma_{DN}=0$).

%%%%%%%%%%%%%%%%%%%%%%%%%%%%%%%%%%%%%%%%%%%%%%%%%%%%%%%%%%%%%%%%%%%%%%%%%%%%5
\begin{figure}[htb]
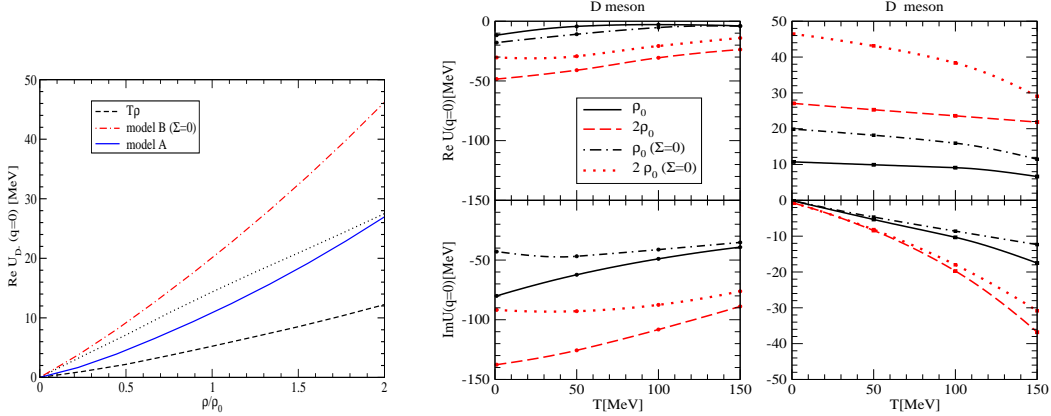

\begin{center}
\includegraphics[height=4.5 cm, width=5 cm]{tolos_fig5.eps}
\hfill
\includegraphics[height=5.5 cm, width=8 cm]{tolos_fig6.eps}
\caption{$\bar D$ mass shift as well as $T\rho$, and the $D$ and $\bar D$ potentials.} \label{fig:DbarN}
\end{center}
\end{figure}
%%%%%%%%%%%%%%%%%%%%%%%%%%%%%%%%%%%%%%%%%%%%%%%%%%%%%%%%%%%%%%%%%%%%%%%%%%%%%

Medium effects at $T=0$ lower the
position of the $\Lambda_c$ and $\Sigma_c$ resonances with respect
to their free  values, in particular with the inclusion of MFB. Their width values, which increase due to $\tilde
Y_c N \rightarrow \pi N \Lambda_c, \pi N \Sigma_c$  processes, differ according to the phase space available. The PD induces a small effect in the resonances because of charm-exchange channels being suppressed, while models A and B are qualitatively similar. Finite temperature results in the reduction of the Pauli blocking effects due to the smearing of the Fermi surface  with temperature. Both resonances move up in energy closer to their free position while they are smoothen out, as in \cite{TOL06}.

In the r.h.s of Fig.~\ref{fig:DN} we display the $D$ meson spectral function for (i) to (iii) (thick lines) for model A and  case (iii) for model B (thin line). At $T=0$ the spectral function presents two peaks: $\tilde \Lambda_c N^{-1}$ excitation at a lower energy whereas the second one
at higher energy is the quasi(D)-particle  peak  mixed with  the $\tilde \Sigma_c N^{-1}$ state. Once MFB is included, the lower peak built up by the $\tilde \Lambda_c N^{-1}$ mode goes up by about $50$ MeV relative to (i) since the meson requires to carry more energy to compensate for the attraction felt by the nucleon.  The same characteristic feature is seen for the $\tilde \Sigma_c N^{-1}$ configuration that mixes with the quasiparticle peak. The PD does not alter much the position of $\tilde \Lambda_c
N^{-1}$ excitation or the quasiparticle peak. For model B ((iii) only), the absence of the $\Sigma_{DN}$ term moves the $\tilde \Lambda_c N^{-1}$ excitation closer to the  quasiparticle peak, while the latter fully mixes with the $\tilde \Sigma_c N^{-1}$ excitation. When finite temperature effects are included, those structures get diluted with increasing temperature while the quasiparticle peak gets closer to its free value and it becomes narrower, because the self-energy
receives contributions from higher momentum $DN$ pairs where the interaction is weaker.

In the $\bar D N$ sector, the scattering lengths for model A (B) are $a^{I=0}=0.61 \ (0)$ fm and   $a^{I=1}=-0.26 \ (-0.29)$ fm.  While our repulsive $I=1$ is 
in good agreement with \cite{LUT06}, the finite
value for the $I=0$ scattering length found in this latter reference is in
contrast to the zero value found here for model B due to the vanishing
$I=0$ coupling coefficient of the corresponding pure TW  $\bar
DN$ interaction. Our results are, however, consistent
with a recent calculation \cite{HAI07}. For model A, we obtain a non-zero value of the $I=0$ scattering length,
due to the magnitude of the $\Sigma_{DN}$ term. As seen in the l.h.s of Fig.~\ref{fig:DbarN}, the $\bar D$ mass shift in cold nuclear matter is repulsive and, in spite of
the absence of resonances close
to threshold, the low-density approximation or $T \rho$ breaks down at relatively low densities. 

Finally, in the r.h.s of  Fig.~\ref{fig:DbarN} we compare the $D$ and
$\bar D$ optical potentials. For model A (B) 
at $T=0$, we obtain an attractive potential of $-12$ ($-18$) MeV for
$D$ meson while the repulsion for $\bar D$ is $11$ ($20$) MeV. A
similar shift in the mass for $D$ mesons is obtained in
Ref.~\cite{TOL06}. The temperature dependence of the repulsive real part of the ${\bar D}$ optical
potential is very weak, while the imaginary part increases steadily due to the
increase of collisional width. The picture is somewhat different for the $D$
meson due to the overlap of the quasiparticle peak with the $\tilde{\Sigma}_c N^{-1}$ mode. The $\tilde{\Sigma}_c N^{-1}$ mode also alters the effect of the $\Sigma_{DN}$ term on the potential.

With regard to the $J/\Psi$ suppression, the in-medium $\bar D$ mass is seen to  increase by about $10-20$ MeV whereas the tail of the quasiparticle peak of the $D$ spectral function extends to lower "mass" due to the thermally spread $\tilde Y_c N^{-1}$. But it is unlikely that this lower tail extends as far down by 600 MeV with sufficient strength.  So the only way for the $J/\Psi$
suppression to take place is by cutting its supply from the excited charmonia: $\chi_{c\ell}(1P)$ or $\Psi'$ by their hadronic collisions, which appears fairly likely kinematically  even at  finite  temperature in the present study.

\end{document}